\documentclass[useAMS,usenatbib,usegraphicx]{mn2e}

 \usepackage{times}

\newcommand{\Mbol}{M _{\rm bol}}

\newcommand{\Teff}{T _{\rm eff}}
\newcommand{\Ks}{K _{\rm s}}
\newcommand{\asterisk}{^{*}}

\title[PLR for type II Cepheids]
{The period-luminosity relation for type II Cepheids
in globular clusters}
\author[N. Matsunaga et al.]
{Noriyuki Matsunaga$^{1}$\thanks{E-mail:matsunaga@ioa.s.u-tokyo.ac.jp},
 Hinako Fukushi$^{1}$,
 Yoshikazu Nakada$^{2,1}$,
 Toshihiko Tanab\'e$^{1}$,
 \newauthor Michael W. Feast$^{3}$,
 John W. Menzies$^{4}$, 
 Yoshifusa Ita$^{5}$, Shogo Nishiyama$^{6}$,
 \newauthor Daisuke Baba$^{6}$,
 Takahiro Naoi$^{5}$, Hidehiko Nakaya$^{7}$, Takahiro Kawadu$^{8}$,
 \newauthor  Akika Ishihara$^{9,10}$ and Daisuke Kato$^{6}$ \\
$^{1}$ Institute of Astronomy, School of Science, the University of Tokyo, 2-21-1 Osawa, Mitaka, Tokyo 181-0015, Japan \\
$^{2}$ Kiso Observatory, Institute of Astronomy, School of Science, the University of Tokyo, Mitake, Kiso, Nagano 397-0101, Japan \\
$^{3}$ Astronomy Department, University of Cape Town, 7701, Rondebosch, South Africa \\
$^{4}$ South African Astronomical Observatory, PO Box 9, 7935, Observatory,
South Africa\\ 
$^{5}$ Institute of Space and Astronomical Science, Japan Aerospace Exploration Agency, \\
Yoshinodai 3-1-1, Sagamihara, Kanagawa 229-8510, Japan \\
$^{6}$ Department of Astrophysics, Nagoya University, Furo-cho, Chikusa-ku, Nagoya 464-8602, Japan \\
$^{7}$ Subaru Telescope, National Astronomical Observatory of Japan, 650 North Aohoku Place, Hilo, HI 96720, USA \\
$^{8}$ Department of Astronomy, Kyoto University, Kitashirakawa-Oiwake-cho, Sakyo-ku, Kyoto 606-8502, Japan \\
$^{9}$ National Astronomical Observatory of Japan, 2-21-1 Osawa, Mitaka, Tokyo 181-8588, Japan \\
$^{10}$ Department of Earth and Planetary Science, School of Science, the University of Tokyo, 7-3-1 Hongo, Bunkyo-ku, Tokyo 113-0033, Japan
}

\begin{document}

\date{Accepted 2006 May 26. Received 2006 May 16;
in original form 2006 March 30}

\pagerange{\pageref{firstpage}--\pageref{lastpage}} \pubyear{2006}

\maketitle

\label{firstpage}

\begin{abstract}
We report the result of our near-infrared observations ($JH\Ks$)
for type II Cepheids (including possible RV Tau stars)
in galactic globular clusters.
We detected variations of 46 variables in 26 clusters
(10 new discoveries in seven clusters) and present their light curves.
Their periods range from 1.2 d to over 80 d.
They show a well-defined period-luminosity relation at each wavelength.
Two type II Cepheids
in NGC~6441 also obey the relation
if we assume the horizontal branch stars 
in NGC~6441 are as bright as
those in metal-poor globular clusters in spite of
the high metallicity of the cluster.
This result supports the high luminosity which has been suggested for the
RR Lyr variables in this cluster. The period-luminosity relation
can be reproduced using the pulsation equation ($P\sqrt{\rho}=Q$)
assuming that all the stars have the same mass.
Cluster RR Lyr variables were found to lie on an extrapolation
of the period-luminosity relation.
These results provide important constraints on the parameters of
the variable stars.

Using Two Micron All-Sky Survey (2MASS) data, we show that
the type II Cepheids in the Large Magellanic Cloud (LMC)
fit our period-luminosity relation within the 
expected scatter at the shorter periods.
However, at long periods ($P>40$ d, i.e. in the RV Tau star range) the
LMC field variables are brighter by about one magnitude 
than those of similar periods in galactic globular clusters.
The long-period cluster stars also differ from both these LMC stars and
galactic field RV Tau stars in a colour-colour diagram. The reasons
for these differences are discussed. 
\end{abstract}


\section{Introduction}
\label{sec:Intro}

Type II Cepheids (hereafter T2Cs) are variables
in the Cepheid instability strip,
but belong to older populations than classical Cepheids.
(see \citealt{Wallerstein-2002}, and references therein, for a review).
They reside in globular clusters, the thick disc, the bulge and the halo,
but not in the thin disc or spiral arms.
Based on their periods, they are often separated into
BL Her stars ($P< 7$ d), W Vir stars ($ 7< P< 20$ d) and
RV Tau stars ($P> 20$ d). The main feature of RV Tau stars is
alternating deep and shallow minima. However, the classification and the nature
of RV Tau stars are ambiguous. Several authors suggested so-called RV Tau stars
include some heterogeneous types of variables.
Whilst six objects in globular clusters have been claimed to be RV Tau stars,
some authors doubted this classification from both the photometric
\citep{Zsoldos-1998} and the spectroscopic point of view
\citep{Russell-1998}. 
In this paper, we will not make a strict distinction between 
RV Tau stars and other T2Cs in clusters.

From previous studies of T2Cs in globular clusters, it is known that they
obey period-luminosity relation (PLR) in the visible ($BVI$).
\citet{Harris-1985} and \citet{McNamara-1995}
claimed the slope of the PLR steepens for periods longer
than about $\log P= 1$.
On the other hand, \citet{Pritzl-2003} did not find such a feature
for the variables in the globular clusters NGC~6388 and NGC~6441.
As \citet{Pritzl-2003} noted, many studies of the T2Cs were
based on old photographic data,
and we need more investigations with modern CCD photometry.
In the near-infrared, no studies have so far been reported.

Studies of variable stars in the near-infrared have become more
numerous in recent times.
For example, many papers have been published on the infrared properties
of RR Lyr variables
(e.g. \citealt{Clement-2001}; Castellani, Caputo \& Castellani 2003).
RR Lyr variables also lie in the Cepheid instability strip but are fainter
than T2Cs.
One of the important motivations for studies of RR Lyr variables is
their application as distance indicators. 
Whilst a larger number of investigations have been devoted to their absolute
visible magnitudes, studies in the infrared have some advantages.
Longmore, Fernley \& Jameson (1986) and Longmore et al. (1990) discovered a
well-defined PLR in the near-infrared for the first time.
It was suggested that the near-infrared relation is less affected by
metal abundance than the visible one, making the near-infrared one
a promising distance indicator.
This led to further works
(\citealt{Butler-2003};
\citealt{Dallora-2004}; \citealt{Storm-2004}; \citealt{DelPrincipe-2005}).
Extensive theoretical studies of the RR Lyr PLR have been also 
carried out (Bono et al. 2001, 2003; Catelan, Pritzl \& Smith 2004;
Di Criscienzo, Marconi \& Caputo 2004).

In this paper, we report the result of our near-infrared observations
for T2Cs in globular clusters and present
their PLR in $JH\Ks$ filters. We also compare the PLR with
that of RR Lyr variables and that of candidate T2Cs
in the Large Magellanic Cloud (LMC).

\section{Observations and Results}
\label{sec:Results}

\subsection{Observations}
Data for T2Cs were obtained during our project to observe
variables of various types in globular clusters. We used
the Infrared Survey Facility (IRSF) 1.4~m telescope and the
Simultaneous 3 Colour Infrared Imager for Unbiused Survey (SIRIUS)
constructed by Nagoya University and the National Astronomical
Observatory of Japan, and sited at the Sutherland station of
the South African Astronomical Observatory. 
Images of a
$7.7 \times 7.7$ arcmin$^2$ field of view are obtained simultaneously
in $JH\Ks$.
The seeing size was typically 1.5 arcsec.
For details of the IRSF and SIRIUS, see \citet{Nagashima-1999}
and \citet{Nagayama-2003}.

Our main targets in the project were red variables with long periods
(100~d or more) \citep{Matsunaga-2006}, so that we observed
each globular cluster only once at a night.
Generally, the clusters were observed once a month between April and
August each year from 2002 to 2005 and on some additional occasions.
The 15 or more observations obtained over this long period enable us
to investigate basic properties of T2Cs.
The survey targeted 145 clusters located south of
about $+30\degr$ Declination.

\subsection{Photometry and Variability detection}

The raw data were reduced in the following way.
We obtained scientific images in $JH\Ks$ filters
for each night using pipeline software (Y Nakajima, private
communication). This involved
dark subtraction, flat-fielding, elimination of hot pixels,
and combination of dithered
images.

For each filter, one of the best images
(weather condition and seeing) was selected as a reference frame
among $N$ images from the repeated observations for a globular cluster.
Photometry was performed on $N$ images with {\small DOPHOT} software
(Schechter, Mateo \& Saha 1993).
In order to standardize the magnitudes, we compared the photometric results
of the reference frames with the Two Micron All-Sky Survey (2MASS) point 
source catalogue
\citep{Curti-2003}.
We found no effect of a colour term  and a typical standard deviation
of about $\pm 0.1$ mag in the difference between 
the magnitudes in the 2MASS catalogue and ours. 
Colour terms were thus  ignored and a constant was added to fit our 
instrumental
magnitudes to those in the 2MASS catalogue.
Note that we could use a large number of objects (say 200 or more)
in these comparisons
so the mean difference of our final magnitudes from the 2MASS system
will be small ( $< 0.01$ mag ).

\begin{figure}
\begin{center}
\includegraphics[clip,width=0.99\hsize]{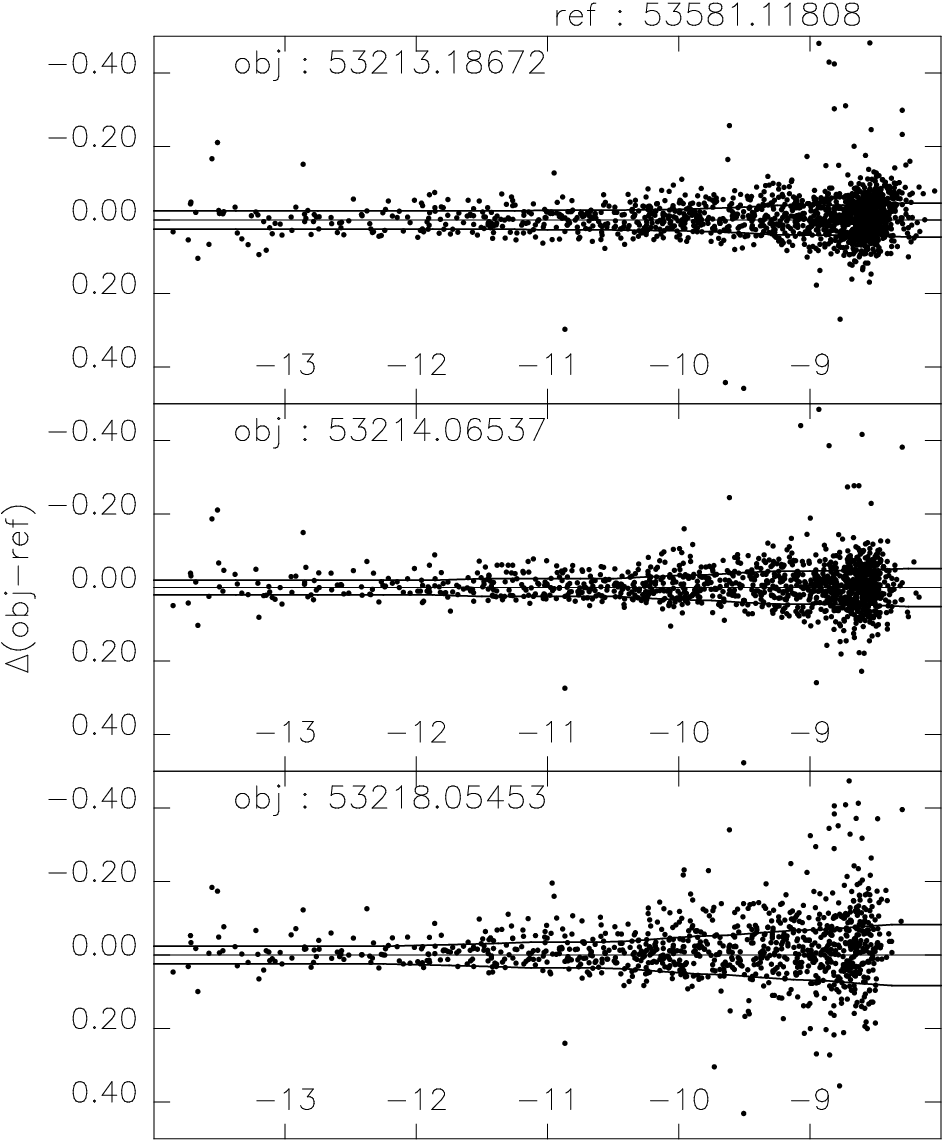}
\caption{An example of the magnitude comparisons between the reference data
(MJD$=53581.11808$) and three repeated observations. These data are
for the cluster NGC 104 (47~Tuc). Magnitudes on the $x$-axis are
instrumental ones before the standardization. Solid curves indicate
the size of the error in each magnitude range.
See the text for details of the analysis.
\label{fig:PhotoComp}}
\end{center}
\end{figure}

The photometric results for the remaining $N-1$ images were compared with 
those of the reference frame,
and we collected differences for all the detected objects.
We present examples of these comparisons in Fig.
\ref{fig:PhotoComp}. 
Variable stars stand out from the general scatter in these plots.
We estimated photometric errors 
as a function of magnitude by taking standard deviations
in boxes of size 0.25~mag or of larger size to
include at least 50 objects and by smoothing the deviations.
The sizes of the estimated errors ($\pm 1\sigma$) are drawn
as solid curves in Fig. \ref{fig:PhotoComp}.
Since some stars have larger errors due to special conditions,
such as being in a crowded region, we adopted the errors from the
{\small DOPHOT} software output if they exceeded the errors just discussed.
We adopted a 3-$\sigma$ cut to distinguish between variable
and non-variable stars.
Celestial coordinates of any detected variable were determined by fitting
to stars in the 2MASS catalogue. The astrometrical precision is expected 
to be
better than 0.5 arcsec in most cases. 

\subsection{Period Determination and the selection condition}
\begin{table}
\begin{minipage}{86mm}
\caption{The objects with periods which are different from
those previously published.
\label{tab:diffP}}
\begin{center}
\begin{tabular}{cccc}
\hline
Cluster & ID & $P$(this work) & $P$(previous) \\
\hline
NGC~5904 & V84 & 26.87 & 26.42 \\
NGC~6218 & V1  & 15.48 & 15.527 \\
NGC~6441 & V6  & 22.47 & 21.365 \\
NGC~7078 & V86 & 16.80 & 17.109 \\
NGC~7089 & V6 & 19.36 & 19.30 \\
\hline
\end{tabular}
\end{center}
\end{minipage}
\end{table}
For any object whose variation was detected in our analysis,
we applied the phase dispersion minimization method
in order to determine a pulsation period \citep{Stellingwerf-1978}.
Even in the case of the long period stars 
(possible RV Tau stars), we did not discriminate between possible
deep and shallow minima since
the number of the minima around which we observed was not large
and the differences in the infrared are small.
We discuss as T2Cs 
in this paper, variables with the following characteristics:\\
~(i) the light curve shows clear periodicity of $1 < P{\rm (d)} < 100$ and\\
~(ii) the location in the colour-magnitude diagram ($J-\Ks$ versus $\Ks$)
is bluer than the red giant branch.\\
In addition, we include in the discussion known T2Cs whose variations
were detected even if our data are not sufficient to determine the periods.
For most of the T2Cs, periods are well determined
from our data, and those obtained from the data in the three filters agree
with each other.
In the case of previously known T2Cs our
periods are consistent with the earlier results
(see \citealt{Clement-2001}).
The earlier optical periods are often  
based on better sampled data
than ours, so that we generally adopt published periods.
For five objects listed in Table \ref{tab:diffP}, however,
our data differ from the previous periods
and we adopted our own values. All five objects
have relatively long periods (W Vir stars or RV Tau stars).
Clement, Hogg \& Yee (1988) pointed out that some of these objects
show rather random changes of the period.
It is known that BL Her stars have rather systematic changes of period
\citep{Wehlau-1982}. However, our observations were not optimized 
to study such effects.

\subsection{Table of T2Cs}

\begin{figure*}
\begin{center}
\includegraphics[clip,width=0.99\hsize]{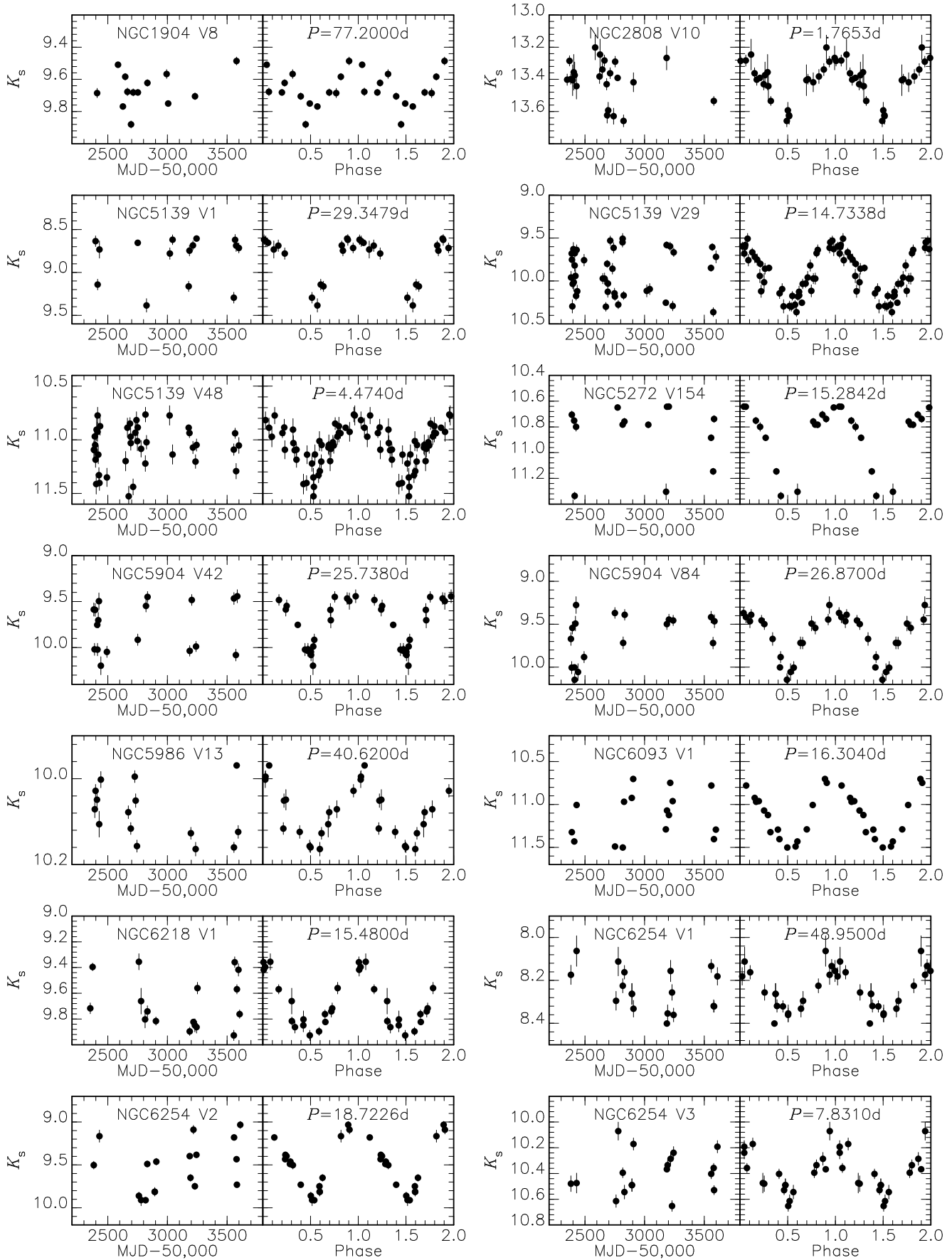}
\caption{Light curves in $\Ks$. Right: plotted against MJDs.
Left: plotted against phases folded by the period indicated
at the top of right panel.\label{fig:LCs}}
\end{center}
\end{figure*}
\addtocounter{figure}{-1}
\begin{figure*}
\begin{center}
\includegraphics[clip,width=0.99\hsize]{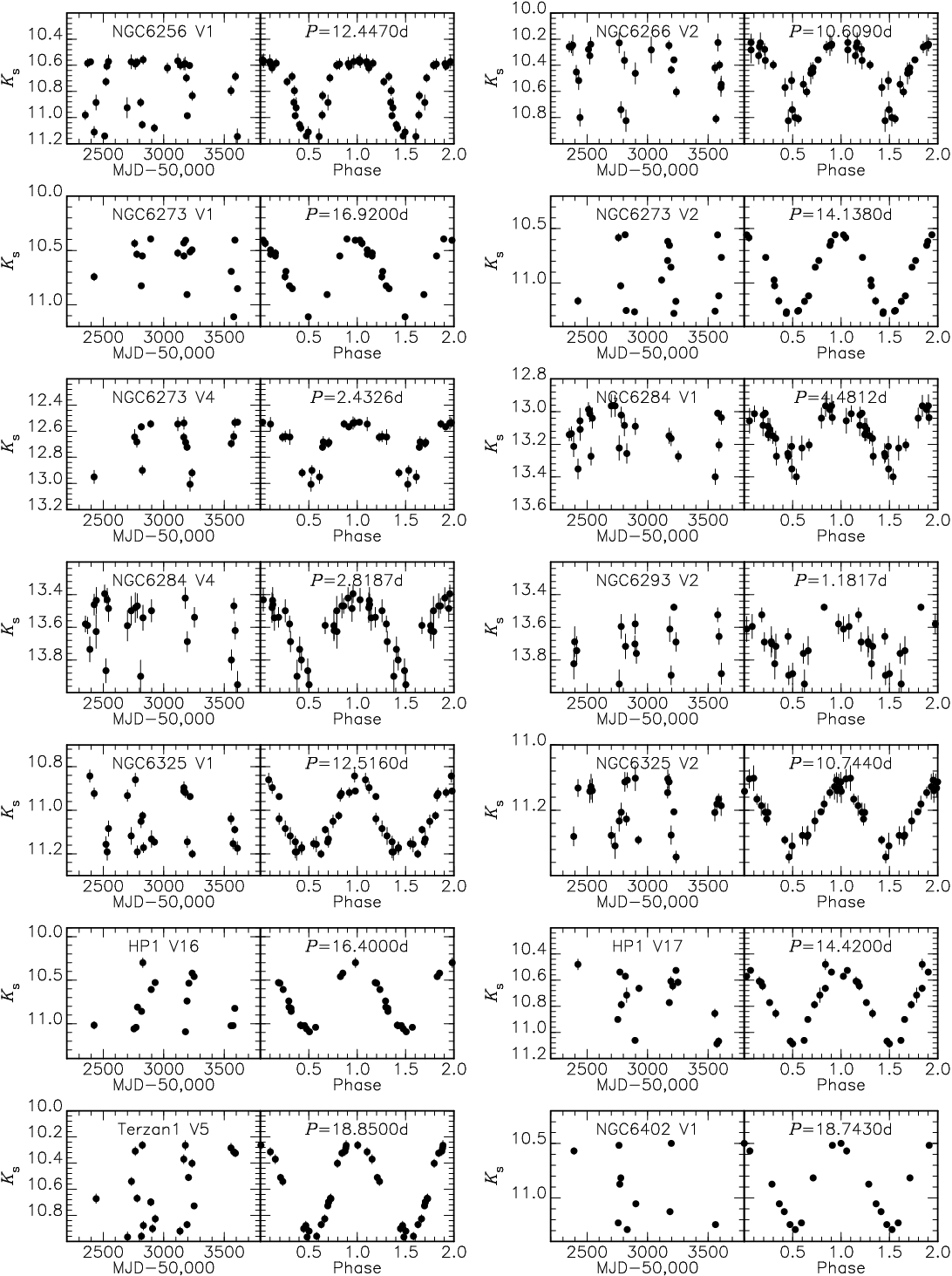}
\caption{-- continued.}
\end{center}
\end{figure*}
\addtocounter{figure}{-1}
\begin{figure*}
\begin{center}
\includegraphics[clip,width=0.99\hsize]{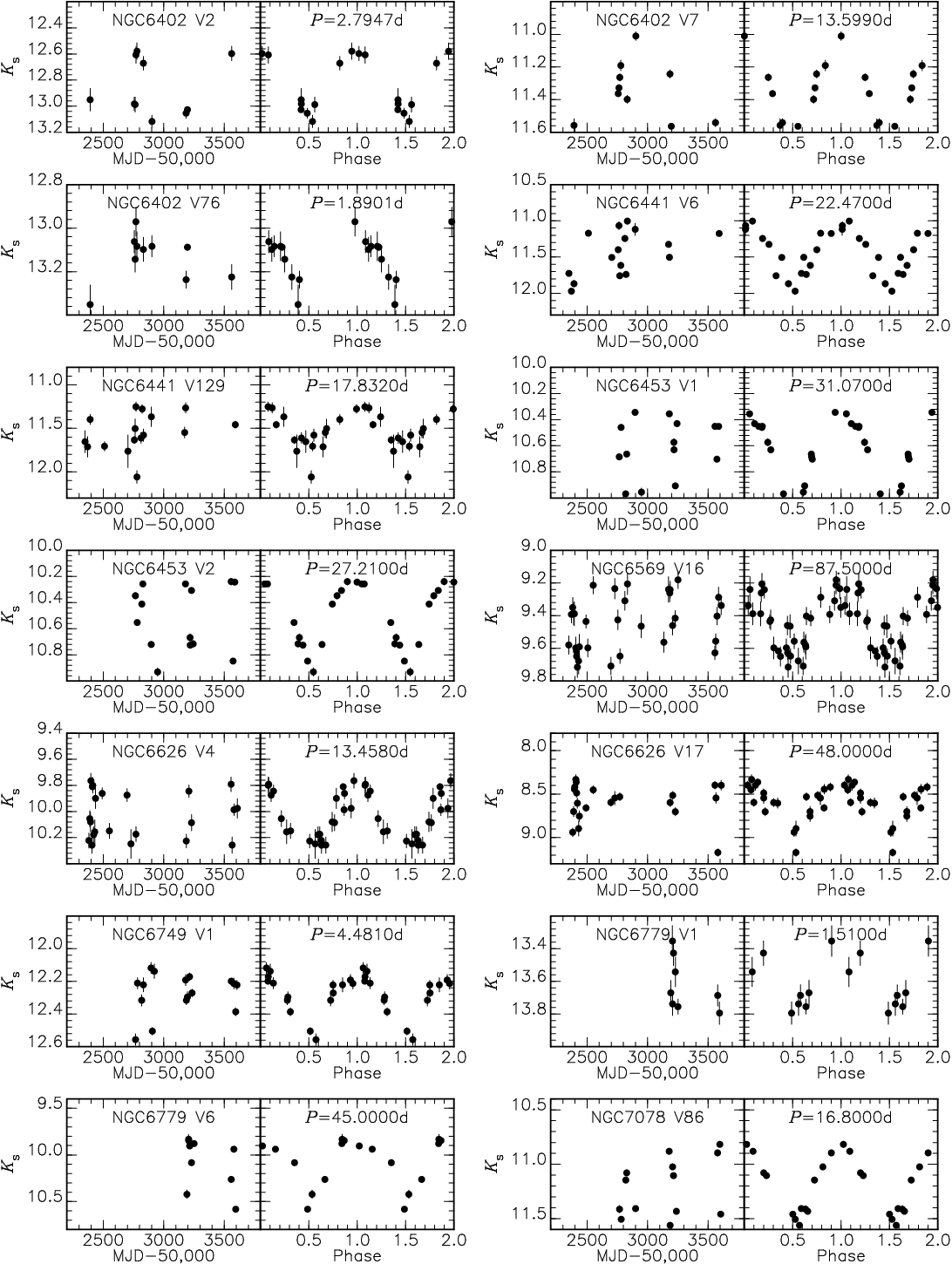}
\caption{-- continued.}
\end{center}
\end{figure*}
\addtocounter{figure}{-1}
\begin{figure*}
\begin{center}
\includegraphics[clip,width=0.99\hsize]{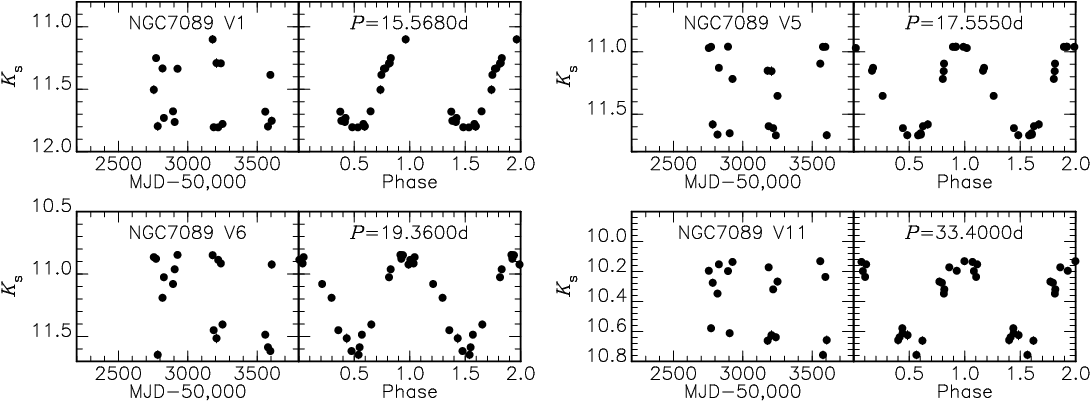}
\caption{-- continued.}
\end{center}
\end{figure*}

\begin{table*}
\begin{minipage}{185mm}
\begin{center}
\caption{List of T2Cs in globular clusters.
$P$ shows a period, $\phi _0$ the date of phase zero
(see the text, for more detail),
$<J>$ to $<K>$ mean magnitudes, 
$\Delta J$ to $\Delta K$ amplitudes,
and $N_{\rm obs}$ the number of observations.
$N_{\rm obs}$ with the superscript $\asterisk$ indicates that
some of the measurements were unavailable.
A flag `n' indicates the object is newly discovered.
\label{tab:Cepheids}}
\begin{tabular}{ccccrcrrrrrrll}
\hline
Cluster & ID & RA(J2000) & Dec(J2000) & $P$ & $\phi _0$ & $<J>$ & $<H>$ & $<\Ks>$ & $\Delta J$ &  $\Delta H$ &  $\Delta \Ks$ & $N_{\rm obs}$ & Flag \\
\hline
NGC1904 &   V8 &  05:24:11.6 & $-$24:31:38 &    77.2 & 53046.168 & 10.36 &  9.84 &  9.68 & 0.38 & 0.40 & 0.39 & 13 &  \\
NGC2808 &  V10 &  09:11:56.9 & $-$64:53:23 &  1.7653 & 53001.125 & 13.89 & 13.54 & 13.43 & 0.45 & 0.35 & 0.46 & 23 $\asterisk$ &  \\
NGC5139 &   V1 &  13:26:05.2 & $-$47:23:43 & 29.3479 & 53013.358 &  9.40 &  9.05 &  8.99 & 0.81 & 0.73 & 0.78 & 15 &  \\
NGC5139 &  V29 &  13:26:27.2 & $-$47:28:48 & 14.7338 & 53007.097 & 10.43 & 10.03 &  9.93 & 0.78 & 0.82 & 0.85 & 40 &  \\
NGC5139 &  V48 &  13:26:37.8 & $-$47:30:25 &   4.474 & 53000.367 & 11.59 & 11.14 & 11.15 & 0.60 & 0.62 & 0.76 & 40 $\asterisk$ &  \\
NGC5272 & V154 &  13:42:11.6 & $+$28:22:14 & 15.2842 & 53003.394 & 11.45 & 11.06 & 10.99 & 0.77 & 0.65 & 0.69 & 14 &  \\
NGC5904 &  V42 &  15:18:24.8 & $+$02:02:53 &  25.738 & 53020.219 & 10.16 &  9.85 &  9.82 & 0.70 & 0.66 & 0.76 & 18 &  \\
NGC5904 &  V84 &  15:18:36.2 & $+$02:04:16 &   26.87 & 53018.678 & 10.20 &  9.80 &  9.71 & 1.08 & 0.95 & 0.87 & 18 &  \\
NGC5986 &  V13 &  15:46:00.3 & $-$37:48:23 &   40.62 & 53009.380 & 10.90 & 10.22 & 10.07 & 0.23 & 0.21 & 0.20 & 15 & n \\
NGC6093 &   V1 &  16:17:04.2 & $-$22:58:54 &  16.304 & 53004.337 & 11.65 & 11.23 & 11.10 & 0.83 & 0.78 & 0.80 & 16 &  \\
NGC6218 &   V1 &  16:47:16.7 & $-$01:57:59 &   15.48 & 53007.412 & 10.24 &  9.79 &  9.64 & 0.64 & 0.61 & 0.57 & 20 $\asterisk$ &  \\
NGC6254 &   V1 &  16:57:10.1 & $-$04:05:36 &   48.95 & 53022.311 &  9.07 &  8.42 &  8.23 & 0.34 & 0.33 & 0.34 & 18 $\asterisk$ &  \\
NGC6254 &   V2 &  16:57:11.7 & $-$04:04:00 & 18.7226 & 53013.656 & 10.05 &  9.61 &  9.47 & 0.90 & 0.94 & 0.88 & 18 $\asterisk$ &  \\
NGC6254 &   V3 &  16:56:56.0 & $-$04:04:16 &   7.831 & 53007.484 & 11.02 & 10.55 & 10.36 & 0.45 & 0.50 & 0.58 & 18 $\asterisk$ &  \\
NGC6256 &   V1 &  16:59:35.0 & $-$37:07:23 &  12.447 & 53004.719 & 11.86 & 11.15 & 10.85 & 0.73 & 0.60 & 0.59 & 30 $\asterisk$ & n \\
NGC6266 &   V2 &  17:01:11.0 & $-$30:07:59 &  10.609 & 53009.157 & 11.22 & 10.64 & 10.53 & 0.76 & 0.63 & 0.60 & 28 $\asterisk$ &  \\
NGC6273 &   V1 &  17:02:38.2 & $-$26:15:12 &   16.92 & 53013.006 & 11.37 & 10.88 & 10.75 & 0.76 & 0.72 & 0.71 & 17 &  \\
NGC6273 &   V2 &  17:02:38.9 & $-$26:13:57 &  14.138 & 53013.840 & 11.53 & 11.06 & 10.92 & 0.82 & 0.74 & 0.72 & 17 &  \\
NGC6273 &   V4 &  17:02:37.6 & $-$26:16:32 &  2.4326 & 53000.138 & 13.28 & 12.85 & 12.77 & 0.49 & 0.47 & 0.47 & 17 &  \\
NGC6284 &   V1 &  17:04:26.9 & $-$24:45:22 &  4.4812 & 53001.750 & 13.68 & 13.24 & 13.18 & 0.38 & 0.41 & 0.43 & 24 &  \\
NGC6284 &   V4 &  17:04:30.3 & $-$24:46:14 &  2.8187 & 53000.533 & 14.15 & 13.71 & 13.67 & 0.64 & 0.52 & 0.56 & 24 &  \\
NGC6293 &   V2 &  17:09:59.8 & $-$26:33:56 &  1.1817 & 53000.392 & 14.26 & 13.81 & 13.71 & 0.34 & 0.25 & 0.47 & 16 &  \\
NGC6325 &   V1 &  17:18:02.5 & $-$23:45:45 &  12.516 & 53003.662 & 11.97 & 11.25 & 11.02 & 0.34 & 0.36 & 0.36 & 24 $\asterisk$ & n \\
NGC6325 &   V2 &  17:17:57.8 & $-$23:46:36 &  10.744 & 53006.200 & 12.14 & 11.43 & 11.22 & 0.24 & 0.26 & 0.24 & 24 & n \\
    HP1 &  V16 &  17:31:08.7 & $-$30:00:22 &    16.4 & 53008.704 & 11.77 & 10.99 & 10.70 & 0.85 & 0.81 & 0.79 & 16 & n \\
    HP1 &  V17 &  17:31:05.7 & $-$29:59:26 &   14.42 & 53004.212 & 11.91 & 11.09 & 10.78 & 0.67 & 0.59 & 0.61 & 16 & n \\
Terzan1 &   V5 &  17:35:46.1 & $-$30:29:03 &   18.85 & 53012.467 & 11.97 & 10.93 & 10.61 & 0.78 & 0.72 & 0.70 & 22 & n \\
NGC6402 &   V1 &  17:37:37.4 & $-$03:14:00 &  18.743 & 53009.449 & 11.63 & 11.10 & 10.89 & 0.85 & 0.80 & 0.79 & 10 &  \\
NGC6402 &   V2 &  17:37:28.6 & $-$03:16:45 &  2.7947 & 53000.084 & 13.45 & 12.98 & 12.85 & 0.52 & 0.52 & 0.54 & 10 &  \\
NGC6402 &   V7 &  17:37:40.4 & $-$03:16:21 &  13.599 & 53012.541 & 12.04 & 11.46 & 11.29 & 0.62 & 0.56 & 0.55 & 10 &  \\
NGC6402 &  V76 &  17:37:29.3 & $-$03:14:45 &  1.8901 & 53001.785 & 13.78 & 13.30 & 13.16 & 0.36 & 0.30 & 0.38 & 10 &  \\
NGC6441 &   V6 &  17:50:15.6 & $-$37:02:16 &   22.47 & 53010.927 & 12.16 & 11.64 & 11.49 & 0.93 & 0.99 & 0.97 & 16 &  \\
NGC6441 & V129 &  17:50:12.9 & $-$37:03:18 &  17.832 & 53001.523 & 12.14 & 11.61 & 11.65 & 0.50 & 0.80 & 0.81 & 16 $\asterisk$ &  \\
NGC6453 &   V1 &  17:50:52.1 & $-$34:36:05 &   31.07 & 53023.985 & 11.51 & 10.85 & 10.66 & 0.70 & 0.62 & 0.62 & 14 & n \\
NGC6453 &   V2 &  17:50:53.0 & $-$34:35:09 &   27.21 & 53016.521 & 11.35 & 10.75 & 10.59 & 0.76 & 0.67 & 0.69 & 14 & n \\
NGC6569 &  V16 &  18:13:37.7 & $-$31:49:13 &    87.5 & 53081.285 & 10.56 &  9.74 &  9.45 & 0.69 & 0.62 & 0.53 & 34 $\asterisk$ &  \\
NGC6626 &   V4 &  18:24:30.1 & $-$24:51:37 &  13.458 & 53005.162 & 10.78 & 10.18 & 10.01 & 0.68 & 0.52 & 0.49 & 24 $\asterisk$ &  \\
NGC6626 &  V17 &  18:24:35.8 & $-$24:53:16 &      48 & 53028.399 &  9.55 &  8.95 &  8.75 & 0.84 & 0.84 & 0.84 & 24 &  \\
NGC6749 &   V1 &  19:05:20.0 & $+$01:55:57 &   4.481 & 53003.020 & 13.38 & 12.62 & 12.34 & 0.38 & 0.44 & 0.44 & 16 & n \\
NGC6779 &   V1 &  19:16:39.3 & $+$30:12:17 &    1.51 & 53000.190 & 13.99 & 13.66 & 13.57 & 0.40 & 0.35 & 0.44 &  9 $\asterisk$ &  \\
NGC6779 &   V6 &  19:16:35.8 & $+$30:11:39 &      45 & 53033.934 & 10.86 & 10.37 & 10.21 & 0.86 & 0.75 & 0.75 &  9 &  \\
NGC7078 &  V86 &  21:29:59.2 & $+$12:10:07 &    16.8 & 53009.520 & 11.70 & 11.32 & 11.19 & 0.84 & 0.83 & 0.74 & 13 &  \\
NGC7089 &   V1 &  21:33:28.5 & $-$00:47:55 &  15.568 & 53008.360 & 11.93 & 11.54 & 11.45 & 0.76 & 0.73 & 0.70 & 18 &  \\
NGC7089 &   V5 &  21:33:23.8 & $-$00:49:13 &  17.555 & 53000.597 & 11.80 & 11.40 & 11.31 & 0.75 & 0.71 & 0.71 & 18 &  \\
NGC7089 &   V6 &  21:33:27.5 & $-$00:50:00 &   19.36 & 53005.973 & 11.72 & 11.33 & 11.25 & 0.86 & 0.79 & 0.80 & 18 &  \\
NGC7089 &  V11 &  21:33:32.4 & $-$00:49:06 &    33.4 & 53024.763 & 10.87 & 10.53 & 10.44 & 0.63 & 0.58 & 0.62 & 18 &  \\
\hline
\end{tabular}
\end{center}
\end{minipage}
\end{table*}
Table \ref{tab:Cepheids} lists 46 T2Cs obtained in our analyses.
We followed the numbering system of variables by \citet{Clement-2001}
and the updated catalogue at their web
page\footnote{http://www.astro.utoronto.ca/\%7Ecclement/read.html}.
We discovered 10 new variables, and gave them successive numbers
after the ones in the Clement's catalogue.
For each variable, Table \ref{tab:Cepheids} lists
the celestial coordinate (RA and Dec.), the period $P$,
the date of phase zero $\phi _0$, mean magnitudes, amplitudes,
and the number of observations $N_{\rm obs}$.
The flag `n' indicates that the object is newly discovered.
The mean magnitudes are taken from the mean of maximum
and minimum magnitudes and the amplitudes are defined as
the minimum-to-maximum variation.
Table \ref{tab:LCtable} lists the individual observations. Only
the first few observations are shown. The full table is given in the
online version of the paper only. In this table,
99.99 is listed when we failed to obtain the magnitude.
This was usually
because the object was fainter than the limiting magnitude
of the frame which depends on
phase and the condition of the frame.
We put a superscript $^*$ to the number of observations $N_{\rm obs}$
in Table \ref{tab:Cepheids}
for an object with the missing measurement(s).
Fig. \ref{fig:LCs} plots the light curves
in $\Ks$ (against modified Julian Dates on the left-hand side
and against phases
folded according to the periods on the right-hand).
By fitting a sine curve to each light curve in $\Ks$,
we determined phases so that the maximum light
of the fitted sine curve occurs at phase zero.
The value $\phi _0$ listed in Table \ref{tab:Cepheids} is
the first date of phase zero after MJD
53000 (2003 December 27).
\begin{table*}
\begin{minipage}{165mm}
\caption{
The first 15 lines in the released table of light variation.
This is a sample of the full version (862 lines), which will be available
in the online version of this journal. Each line lists the data of
each observation: MJD (modified Julian Date), phase (zero for the maxima),
magnitudes ($JH\Ks$), and errors ($E_J, E_H$ and $E_{\Ks}$).
\label{tab:LCtable}}
\begin{center}
\begin{tabular}{lrcccccccc}
\hline
 Cluster & ID & MJD & Phase & $J$ & $E_J$ & $H$ & $E_H$ & ${\Ks}$ & $E_{\Ks}$ \\
\hline
 NGC1904 &  V8 &52410.6926 &0.688 &10.38 &0.02 & 9.86 &0.02 & 9.69 &0.03 \\
 NGC1904 &  V8 &52586.1033 &0.960 &10.21 &0.03 & 9.67 &0.02 & 9.51 &0.02 \\
 NGC1904 &  V8 &52627.0778 &0.491 &10.45 &0.03 & 9.91 &0.02 & 9.77 &0.01 \\
 NGC1904 &  V8 &52646.0534 &0.736 &10.25 &0.03 & 9.77 &0.02 & 9.58 &0.02 \\
 NGC1904 &  V8 &52665.0233 &0.982 &10.36 &0.02 & 9.86 &0.03 & 9.68 &0.02 \\
 NGC1904 &  V8 &52694.8237 &0.368 &10.55 &0.03 &10.04 &0.02 & 9.88 &0.02 \\
 NGC1904 &  V8 &52713.8921 &0.615 &10.34 &0.02 & 9.86 &0.02 & 9.68 &0.02 \\
 NGC1904 &  V8 &52752.7671 &0.119 &10.38 &0.03 & 9.87 &0.02 & 9.68 &0.02 \\
 NGC1904 &  V8 &52832.2021 &0.148 &10.34 &0.02 & 9.79 &0.02 & 9.62 &0.02 \\
 NGC1904 &  V8 &52993.0750 &0.231 &10.27 &0.02 & 9.75 &0.01 & 9.57 &0.03 \\
 NGC1904 &  V8 &53006.9818 &0.412 &10.46 &0.02 & 9.93 &0.01 & 9.75 &0.01 \\
 NGC1904 &  V8 &53231.1906 &0.316 &10.40 &0.03 & 9.88 &0.01 & 9.71 &0.02 \\
 NGC1904 &  V8 &53579.1520 &0.823 &10.18 &0.02 & 9.64 &0.02 & 9.49 &0.03 \\
 NGC2808 & V10 &52351.8047 &0.702 &13.87 &0.03 &13.48 &0.03 &13.40 &0.04 \\
 NGC2808 & V10 &52370.9252 &0.533 &13.78 &0.04 &13.36 &0.04 &13.29 &0.03 \\
\hline
\end{tabular}
\end{center}
\end{minipage}
\end{table*}

Some of the known variables listed in Table \ref{tab:Cepheids}
with long periods, say $P > 20$ d, were not classified as
T2Cs (or RV Tau stars) in previous work.
For example, NGC 6254 V1 was classified as a semi-regular
type variable by Clement, Hogg \& Wells (1985).
It is difficult to separate the light curves of these stars
from those of red variables.
However, our sample is clearly defined
(see previous section)
and the stars we consider as T2Cs are all bluer in 
$J-\Ks$ than the giant branch of the clusters.

Among about 80 known, or suspected,
T2Cs in clusters, about half are not included in this work.
Some of them were not targetted in our
observations because they are located too far north (Dec.$>30\degr$)
or too far from the cluster centre for the field of view
of our camera. The others are either too faint for useful 
photometry or blended with a neighbouring red giant.

\subsection{Chances of the contamination of field T2Cs\label{sec:density}}

In Section \ref{sec:PLR}, we show that our T2Cs define a narrow PLR.
It is therefore unlikely that any of them are cluster nonmembers.
However it is of interest to make some estimate of the chance of
encountering a field T2C in our survey.

In the General Catalogue of Variable Stars
\citep{Kholopov-1998}, there are 178 variables listed as CW (W Vir and BL Her)
and 126 variables listed as RV (RV Tau).
Considering their distribution over the sky, we divide them into three groups
according to galactic coordinates $(l,b)$:
the halo ($|b|>10\degr$), the bulge ($|l|<10\degr$ and $|b|<10\degr$),
and the disc ($|l|>10\degr$ and $|b|<10\degr$). The number $N$
and the corresponding density $\sigma$ (str$^{-1}$) of the variables
are listed in Table \ref{tab:density} for each group. 
Unfortunately, the list of field T2Cs is not based 
a complete, uniform, survey.
Recently, \citet{Kubiak-2003} presented the result of a
T2C survey with the Optical Gravitational Lensing Experiment (OGLE) data base.
They found 54 T2Cs in about 11 square degrees of the Galactic bulge.
This corresponds to the density of $\sigma=1600$ (str$^{-1}$),
which is larger than the value listed in Table \ref{tab:density}
by a factor of three.
The solid angle of a field of view of the SIRIUS camera is
$5.0\times 10^{-6}$ (str) and we observed
43 globular clusters within the bulge region.
Therefore, the expected number of field T2Cs in our survey is
small, less than 0.4, even in the high field density of the bulge.
Since many of the clusters discussed in this paper
are in much lower density environments than the bulge,
the expected number of field interlopers is much less than this
and can be neglected.

\begin{table}
\begin{minipage}{86mm}
\caption{
The density of T2Cs based on the General Catalogue of Variable
Stars\citep{Kholopov-1998}. The regions are separated by galactic coordinate
(see the text). `CW' and `RV' are the classified types in the catalogue for
the combination of BL Her and W Vir (CW) and for RV Tau (RV).
\label{tab:density}}
\begin{center}
\begin{tabular}{lcccccc}
\hline
Region & \multicolumn{2}{c}{CW} & \multicolumn{2}{c}{RV} &  \multicolumn{2}{c}{CW+RV} \\ 
& $N$ & $\sigma$ & $N$ & $\sigma$ & $N$ & $\sigma$ \\
& & (str$^{-1}$) & & (str$^{-1}$) & & (str$^{-1}$) \\
\hline
Halo & 85 & 8.2 & 47 & 4.5 & 132 & 12.7 \\
Bulge & 53 & 438 & 25 & 207 & 78 & 645 \\
Disk & 40 & 19.4 & 54 & 26.2 & 94 & 45.6 \\
All sky & 178 & 14 & 126 & 10 & 304 & 24 \\
\hline
\end{tabular}
\end{center}
\end{minipage}
\end{table}

\subsection{Parameters for globular clusters}

Now we turn to absolute magnitudes of T2Cs to combine those in
different globular clusters into a period-luminosity diagram.
The distance moduli we adopt are based on the
magnitudes of horizontal branches of the clusters.
We adopted the relation
\begin{equation}
M{\rm _V (HB)} = 0.22 {\rm [Fe/H]} + 0.89
\label{eq:MvHB}
\end{equation}
from \citet{Gratton-2003}, who calibrated the relation by using
the main-sequence fitting method for three clusters.
Relations similar to this have also been derived by others.
We adopted the values listed in the table compiled by \citet{Harris-1996},
for the metal abundance [Fe/H], the apparent magnitude of horizontal branch
$V$(HB), and the reddening $E(B-V)$. 
We used the version released in 2003 February, updated in his web page
\footnote{http://www.physics.mcmaster.ca/Globular.html},
except in the case of HP~1 for which we assumed the values in the version
released in 1997 May (see 
\ref{sec:comments}).
For the reddening corrections, we used $R_{\rm V}=3.1$ and
the following extinction law,
\begin{equation}
\frac{A_J}{E(B-V)} = 0.866,~ 
\frac{A_H}{E(B-V)} = 0.565,~
\frac{A_{\Ks}}{E(B-V)} = 0.365
\label{eq:ExtLaw}
\end{equation}
adopted from Cardelli, Clayton \& Mathis (1989).

It is known that NGC~5139 ($\omega$ Cen) has a metallicity spread and contains
a population as metal rich as [Fe/H] $\sim -0.6$.
However, in this cluster, a large population of horizontal branch stars
and RR Lyr stars belong to metal-poor populations and
the metallicity distribution peaks at around [Fe/H]$=-1.6$
\citep{Sollima-2006}.
Adopting [Fe/H]$=-1.6$, the distance modulus
$(m-M)_0$ is derived to be 13.62, which agrees with 
the value obtained from an eclipsing binary in the cluster\citep{Thompson-2001}.
NGC~6441 is another cluster for which we need to take special care.
It has a peculiar horizontal branch and contains blue HB stars and RR Lyr stars
in spite of its high metallicity \citep{Pritzl-2003}.
We adopted [Fe/H]$=-2.0$ for RR Lyr stars in this cluster
as \citet{Pritzl-2003} did,
and we inserted it into equation (\ref{eq:MvHB}).
We will give more detailed discussion in Section \ref{sec:comments}.
Our adopted metallicities, reddenings, $V$(HB)s and distance
moduli are listed in Table \ref{tab:CLs}.

\begin{table}
\begin{minipage}{86mm}
\begin{center}
\caption{Parameters for globular clusters. The metallicity [Fe/H], the colour excess $E(B-V)$, and the magnitude of the horizontal branch $V$(HB) were adopted from \citet{Harris-1996}. The distance modulus $(m-M)_0$ was estimated from equation (\ref{eq:MvHB}), except for NGC~6441 (see the text).\label{tab:CLs}}
\begin{tabular}{lllll}
\hline
 Cluster  & [Fe/H] & $E(B-V)$ & $V$(HB) & $(m-M)_0$ \\
\hline
NGC~1904 &  $-$1.57 & 0.01 & 16.15 & 15.57 \\
NGC~2808 &  $-$1.15 & 0.22 & 16.22 & 14.90 \\
NGC~5139 &  $-$1.6  & 0.12 & 14.53 & 13.62 \\
NGC~5272 &  $-$1.57 & 0.01 & 15.68 & 15.10 \\
NGC~5904 &  $-$1.27 & 0.03 & 15.07 & 14.37 \\
NGC~5986 &  $-$1.58 & 0.28 & 16.52 & 15.11 \\
NGC~6093 &  $-$1.75 & 0.18 & 16.10 & 15.04 \\
NGC~6218 &  $-$1.48 & 0.19 & 14.60 & 13.45 \\
NGC~6254 &  $-$1.52 & 0.28 & 14.65 & 13.23 \\
NGC~6256 &  $-$0.70 & 1.03 & 18.50 & 14.57 \\
NGC~6266 &  $-$1.29 & 0.47 & 16.25 & 14.19 \\
NGC~6273 &  $-$1.68 & 0.41 & 16.50 & 14.71 \\
NGC~6284 &  $-$1.32 & 0.28 & 17.40 & 15.93 \\
NGC~6293 &  $-$1.92 & 0.41 & 16.50 & 14.76 \\
NGC~6325 &  $-$1.17 & 0.89 & 17.90 & 14.51 \\
NGC~6402 &  $-$1.39 & 0.60 & 17.30 & 14.86 \\
HP~1     &  $-$1.50 & 1.19 & 18.60 & 14.36 \\
Terzan~1 &  $-$1.30 & 2.28 & 21.40 & 13.73 \\
NGC~6441 &  $-$0.53 & 0.47 & 17.51 & 15.60 \\
NGC~6453 &  $-$1.53 & 0.66 & 17.53 & 14.93 \\
NGC~6569 &  $-$0.86 & 0.55 & 17.52 & 15.11 \\
NGC~6626 &  $-$1.45 & 0.40 & 15.55 & 13.74 \\
NGC~6749 &  $-$1.60 & 1.50 & 19.70 & 14.51 \\
NGC~6779 &  $-$1.94 & 0.20 & 16.16 & 15.08 \\
NGC~7078 &  $-$2.26 & 0.10 & 15.83 & 15.13 \\
NGC~7089 &  $-$1.62 & 0.06 & 16.05 & 15.33 \\
\hline
\end{tabular}
\end{center}
\end{minipage}
\end{table}

\subsection{Period-luminosity relation\label{sec:PLR}}

\begin{figure}
\begin{center}
\includegraphics[clip,width=0.99\hsize]{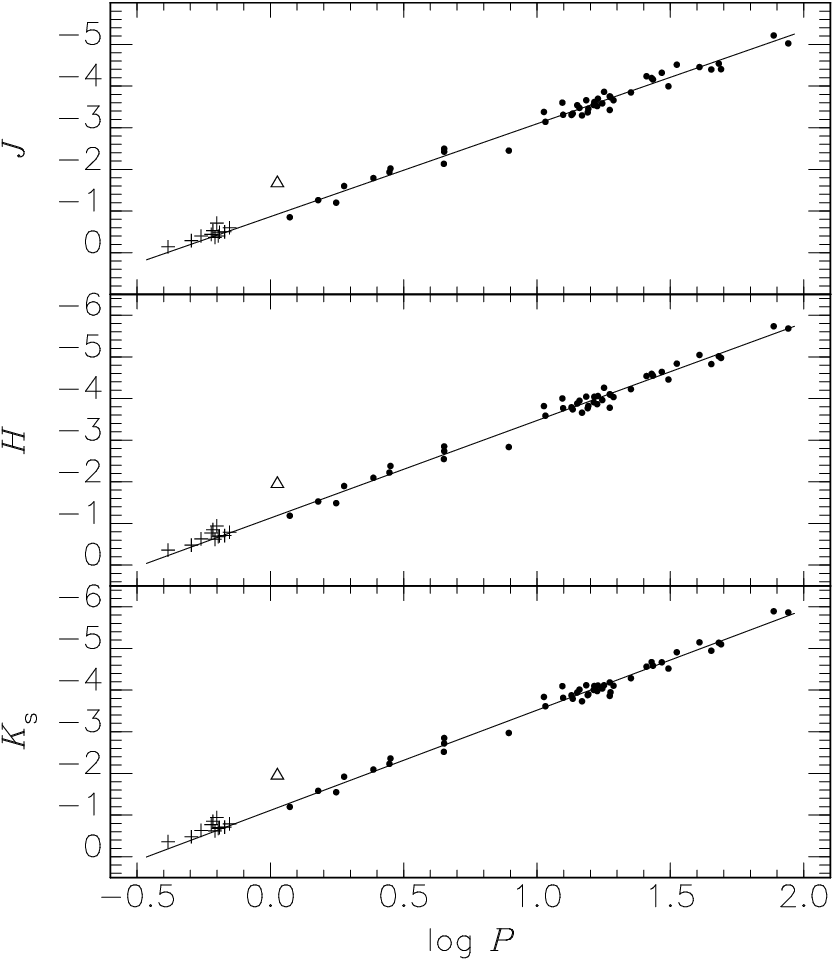}
\caption{
Period-luminosity relation in $JH\Ks$ filters for type II Cepheids
(filled circles). 
Linear regressions (to filled circles) are shown as solid lines.
The data for RR Lyr variables in NGC~6341 are also plotted as plus symbols
taken from \citet{DelPrincipe-2005}.
The triangle at $\log P = 0.026$ is the data for NGC~6341 V7.
See the discussion in Section \ref{sec:RRcomp} for RR Lyr variables and
NGC~6341 V7.
\label{fig:RRcomp}}
\end{center}
\end{figure}

The distance moduli and reddenings discussed in the last section
were used to derive absolute magnitudes,
and we obtained
period-luminosity diagrams in three filters (Fig. \ref{fig:RRcomp}).
Linear regressions to the T2C data (filled circles) yield,
\begin{eqnarray}
M_J &=& -2.23 ~(\pm 0.05) (\log P -1.2) - 3.54 ~(\pm 0.03),
\label{eq:PLR_J} \\
M_H &=& -2.34 ~(\pm 0.05) (\log P -1.2) - 3.94 ~(\pm 0.02),
\label{eq:PLR_H} \\
M_{\Ks} &=& -2.41 ~(\pm 0.05) (\log P -1.2) - 4.00 ~(\pm 0.02),
\label{eq:PLR_K}
\end{eqnarray}
with residual standard deviations of 0.16, 0.15 and 0.14 mag,
respectively.

\citet{Arp-1955} and 
Nemec, Nemec \& Lutz (1994) claimed 
that there were fundamental-mode 
and first-overtone-mode pulsators forming separate parallel sequences
in the T2C PLR.
On the other hand, \citet{McNamara-1995}
doubted the existence of any overtone pulsators.
Fig. \ref{fig:RRcomp} shows there is no evidence for more than one
mode of pulsation.
As already mentioned, some papers claimed that the slope 
of the PL in the optical gets steeper for
T2Cs at around
$\log P=1$ (\citealt{Harris-1985}; \citealt{McNamara-1995}),
whilst \citet{Pritzl-2003} did not find such an effect.
As Fig. \ref{fig:RRcomp} shows there is no evidence for other
than a linear relation in the near-infrared over the whole period range.

\subsection{Comments on some clusters\label{sec:comments}}

\subsubsection{HP~1}
The two T2Cs in HP~1 are of nearly the same magnitudes and periods
strongly suggesting a common distance and making it unlikely
that they are field stars. However, if we use the reddening in the 2003 
version of the Harris catalogue [$E(B-V) = 0.74$ derived by
\citet{Davidge-2000} from infrared observations of field stars],
these stars lie above the PLR by amounts which
depend on wavelength. This wavelength dependence is symptomatic of
an incorrect reddening correction. On the other hand, using the 
reddening from
the 1997 version of the Harris catalogue places the star on the PLR
at all wavelengths. This latter reddening ($E(B-V)= 1.19$) was obtained by
Ortolani, Bica \& Barbuy (1997) from a comparison of 
the $(V-I)$ colour of the RGB with 
that of NGC 6752. Other authors have also found reddenings larger than that
derived by Davidge
(i.e. $E(B-V)=1.44$, \citealt{Armandroff-1988}; 1.88, \citealt{Minniti-1995}).
We have therefore used the Ortolani value of the reddening.

\subsubsection{NGC~6441}
Despite the relatively high metallicity usually adopted for NGC~6441
([Fe/H] $= -0.53$; \citealt{Harris-1996}), 
\citet{Rich-1997} discovered it to have a blue horizontal branch 
as does the related cluster NGC~6388.
These clusters have many RR Lyr variables (\citealt{Layden-1999};
\citealt{Pritzl-2003}, and references therein),
and these RR Lyr stars resemble those in Oosterhoff II clusters (metal poor)
rather than those in Oosterhoff I ones (relatively metal rich).
Nevertheless, \citet{Clementini-2005} have recently reported that
RR Lyr stars in NGC~6441 are not of low metallicity.
\citet{Pritzl-2000} suggested
that they comprise a new Oosterhoff group
and they found circumstantial evidence that they may be at least 
as bright as those in the very metal-poor clusters. 
In view of these results we have followed \citep{Pritzl-2003} and used
an absolute magnitude for the
HB of this cluster equivalent to that of one with [Fe/H] $= -2.0$.
There is very little doubt that the T2Cs belong to NGC~6441 
since there are still four more T2Cs observed optically,
but in the crowded cluster centre not observed by us,
besides the two discussed here. They have (optical) magnitudes
consistent with the ones we have studied \citep{Pritzl-2003}. 
In so much as the T2Cs in NGC~6441 fall on our PL relation at our
adopted distance, they support the high luminosities for the
RR Lyr variables as discussed by \citet{Pritzl-2000}.

\section{Discussion}
\label{sec:Discussion}

\subsection{Metallicity effect on the PLR}

\begin{figure}
\begin{center}
\includegraphics[clip,width=0.99\hsize]{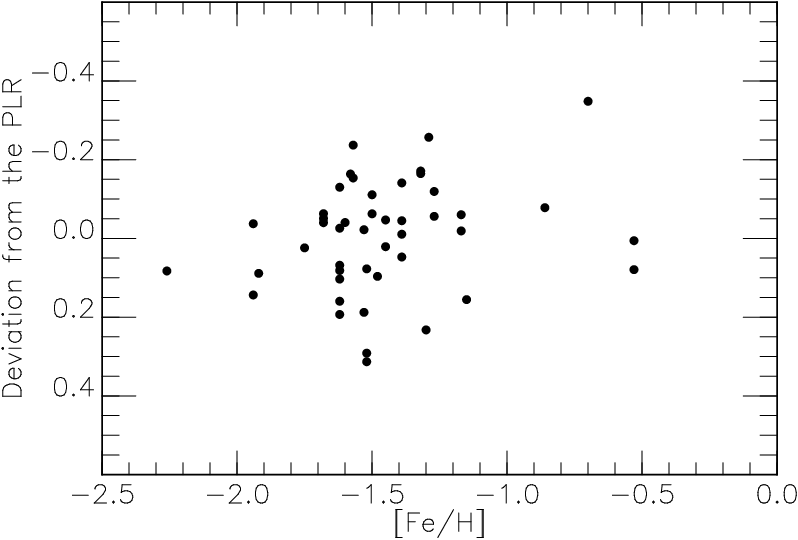}
\caption{
The relation between the metallicity and the deviation from the PLR in $\Ks$.
\label{fig:deltaPLR}}
\end{center}
\end{figure}

First, we discuss the metallicity effect on the zero-point of the PLR,
by comparing deviations
from the PLR (\ref{eq:PLR_J}) -- (\ref{eq:PLR_K})
with the metallicity for each object. We simply adopted the metallicity of
the globular cluster, in which a T2C is found,
as the metallicity of the T2C.
The relation in the $\Ks$ filter is shown in Fig. \ref{fig:deltaPLR}.
The distance moduli we used equation (\ref{eq:MvHB})
have of course a metallicity dependence by themselves. 
A linear regression for the data in Fig. \ref{fig:deltaPLR}
has a slope of $-0.10(\pm 0.06)$ which
is hardly significant. It would be reduced to $-0.02$
if we adopted the slope of 0.30 for equation (\ref{eq:MvHB}) derived earlier by
\citet{Sandage-1993}.
Note that adopting the latter slope makes negligible difference (less than
1 percent) to our PLR slopes.

\subsection{Extension of the PLR to RR Lyr region\label{sec:RRcomp}}

\begin{figure}
\begin{center}
\includegraphics[clip,width=0.99\hsize]{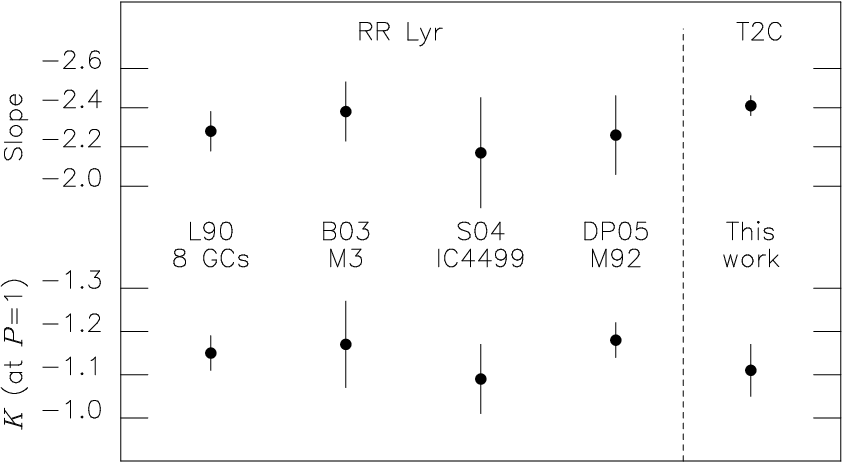}
\caption{
Comparison of the slope and the zero-point of the PLR
of T2Cs (this work) and those of RR Lyr variables in references.
L90 shows the result of \citet{Longmore-1990}, B03 \citet{Butler-2003},
S04 \citet{Storm-2004} and DP05 \citet{DelPrincipe-2005}.
\label{fig:compPLRs}}
\end{center}
\end{figure}

We found that RR Lyr variables also obey the PLR (\ref{eq:PLR_J}) --
(\ref{eq:PLR_K}).
Plus symbols in Fig. \ref{fig:RRcomp} indicate RR Lyr variables in NGC~6341
(M~92) taken from \citet{DelPrincipe-2005}.
We adopted a distance modulus of 14.65~mag obtained in the same manner
as for other clusters (equation \ref{eq:MvHB}). This is the only cluster
with RR Lyr observations at all three wavelengths.
A comparison can be made at $\Ks$ for a number of other clusters.
As shown in Fig. \ref{fig:compPLRs},
both the slope and the zero-point of the PLR agree satisfactorily 
with that of the RR Lyr variables in all
cases.
The data for the RR Lyr variables are from
\citep{Longmore-1990}, \citep{Butler-2003}, \citep{Storm-2004},
and \citep{DelPrincipe-2005}.
We averaged the results for the eight globular clusters
in \citet{Longmore-1990}.
Although their magnitudes are
$K$ (not $\Ks$) in various photometric system,
the differences are negligible (less than 0.01 mag; see
\citealt{Carpenter-2001}, for example).
Theoretical studies also provided sufficiently close PLRs.
For example, the slope in \citet{Bono-2001} was $-$2.07 and that in
\citet{Catelan-2004} was $-$2.35. These works showed that
there is a small metallicity dependence of the zero-point ($\sim 0.17 \log Z$
in $K$), but the effect is not clear
in the observational results and must be small
(\citet{Longmore-1990} derived the 
metallicity-dependent term as $0.04$[Fe/H]).
These results carry the implication that
stars with the same age and probably the same mass
within the instability strip obey the same PLR.
We will discuss this in next section.

Some comments should be provided about NGC~6341 V7 \citep{DelPrincipe-2005},
which has a period of $\log P = 0.0259$ and is deviant
from the PLR (the triangle in Fig. \ref{fig:RRcomp}).
Unfortunately, NGC~6341 lies too north to be observed by us.
\citet{Kopacki-2001} reported that this star is a BL Her star,
but it is apparently brighter than expected from our PLR.
Although we need to confirm its membership,
the location of NGC6341 in the halo indicates
that this star belongs to the cluster (see Section \ref{sec:density}).
In that case, this star may be the second anomalous Cepheid
in globular clusters after NGC~5466 V19.
An anomalous Cepheid is a more massive variable with $0.3 < P({\rm d}) < 2$
and is brighter than RR Lyr variables by about one magnitude
(\citealt{Zinn-1976}; \citealt{Cox-1988}). 
This star needs further investigation, including a confirmation
of cluster membership.
\subsection{Reproduction of the PLR}

Here, we discuss the PLR by using the $P\sqrt{\rho} = Q$ relation.
The relation can be written as
\begin{eqnarray}
\Mbol &=& -3.33 \log P - 1.67 \log M - 10 \log \Teff  \nonumber \\
      & & + M_{\rm bol,\odot} + 10 \log T_{\rm eff,\odot} + 3.33 \log Q,
\label{eq:Pul}
\end{eqnarray}
where $M$ is the mass in units of the solar mass,
and $\Teff$ the effective temperature (e.g. \citealt{McNamara-1995}).
Two linear relations,
\begin{eqnarray}
\log \Teff &=& -0.058 \log P + 3.81 ,
\label{eq:TP} \\
\log Q &=& 0.24 \log P -1.39 ,
\label{eq:QP}
\end{eqnarray}
are adopted from \citet{McNamara-1994} to derive the PLR.
We express the bolometric correction as
\begin{equation}
\Mbol - M_\lambda = \alpha _\lambda \log \Teff + \beta _\lambda .
\label{eq:BCP}
\end{equation}
Using these relations,
eq (\ref{eq:Pul}) can be expressed in the form,
\begin{eqnarray}
M_\lambda &=&
-(1.95-0.058 \alpha _\lambda) (\log P-1.2) -1.67\log {\rm M} \nonumber \\
& & - (2.70 +3.74\alpha _\lambda +\beta _\lambda)
\label{eq:Pul2}
\end{eqnarray}
with the constants of 
$M_{\rm bol,\odot}=4.75$ and $T_{\rm eff,\odot}=5780{\rm K}$. 
We define $\mu _\lambda$ and $\eta _\lambda$ as the dependence
on the period (${\rm d}M_\lambda /{\rm d}\log P$) and the zero-point
of the relation at $\log P=1.2$, respectively
[i.e. $M_\lambda = \mu _\lambda (\log P -1.2) + \eta _\lambda$].
If the mass term ($\log M$) has
no dependence on the period, $\mu _\lambda$ equals to 
$-(1.95-0.058\alpha _\lambda)$.
However, Bono, Caputo \& Santolamazza (1997)
predicted that the mass of T2Cs varies from $0.59M _\odot$
to $0.52M _\odot$, decreasing with increasing period from
1 to 10 d. This period dependence increases $\mu _\lambda$ by
0.08 compared with the case of the constant mass.

We obtained $\alpha _\lambda$ and $\beta _\lambda$ in Table \ref{tab:PLRslope}
from the model atmospheres listed in table 1 of
Bessell, Castelli \& Plez (1998). 
They listed both the models with overshooting (table 1) and those without
overshooting (table2), but the difference between the two sets has
negligible effect on our results (up to 0.03 mag).
Whilst these models are computed with the solar metallicity,
Sandage, Bell \& Tripicco (1999) computed model atmospheres for Cepheids
between [Fe/H]$=0.0$ and $-1.7$. Their results show that
the slopes of the $\log \Teff$-bolometric correction relation (equation \ref{eq:BCP})
are within the uncertainty of our adopted values, and
the zero-points get slightly smaller for the lower metallicity
(about 0.1-mag difference between [Fe/H]$=0.0$ and $-1.7$).
In $JH\Ks$ filters we would expect any effect of metallicity on
equation (\ref{eq:BCP}) to be less than in $V$ and $I$.

\begin{figure}
\begin{center}
\includegraphics[clip,width=0.99\hsize]{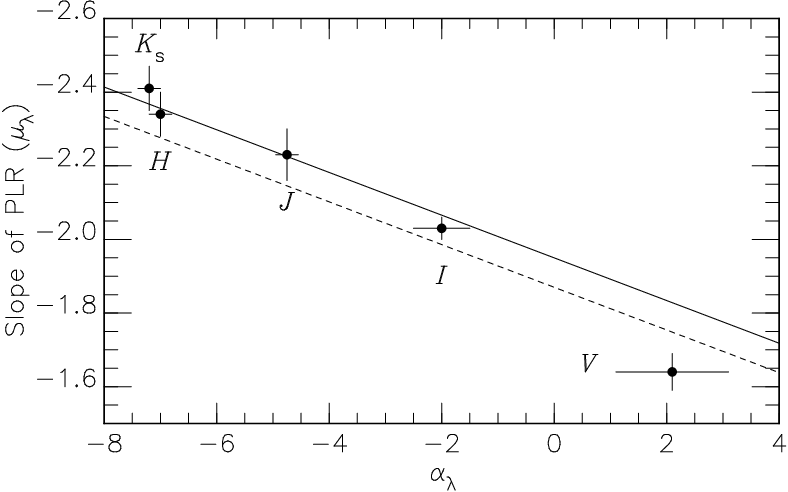}
\caption{The relation between the adopted bolometric correction
($\alpha _\lambda$) and the slope of the PLR ($\mu _\lambda$).
The solid line indicates the relation according to equation (\ref{eq:Pul2})
in case of the constant mass, and the broken one is the one shifted
by 0.08 considering the mass dependency mentioned in the text.
Filled circles show the observational results.
\label{fig:PLRslope}}
\end{center}
\end{figure}

The third and fourth columns in Table \ref{tab:PLRslope} show
the  predicted slope $\mu _\lambda$ and zero-point $\eta _\lambda$
in case of constant mass. They are approximately consistent with
the counterparts obtained from the observational data
(the fifth and sixth columns).
The observational values in $V$ and $I$ filters
are taken from \citet{Pritzl-2003}
and those in $JH\Ks$ filters are obtained by us.
Fig. \ref{fig:PLRslope} shows the relation between $\alpha _\lambda$
and $\mu _\lambda$.
The solid line shows the case of constant mass,
and the broken one shows the one shifted by 0.08 for the mass-dependent case.
The observational values (filled circles) 
favour the constant mass model at least in our own data ($JH\Ks$).
In $V$, the slope expected from equation (\ref{eq:Pul2}) is steeper than
the observational value.
A linear fit to the bolometric correction (equation \ref{eq:BCP})
is not very  good in $V$, since
a quadratic term is no longer small unlike the case of $JH\Ks$.
The value of $\alpha _V$ in equation (\ref{eq:BCP}) can range from 0 to 4
between the extreme values of
$\log \Teff$, $3.85$ and $3.70$. If we adopt a quadratic relation
instead of equation (\ref{eq:BCP}), the PLR also become quadratic
and the slope of a linear fit to the entire period range is about $-1.7$
in $V$, which is close to the observational value ($-1.64$).
Since our discussion is based on very simple scheme, more detailed
work on both the theoretical and observational side is desirable.

It is worth noting that T2Cs and RR Lyr variables may be unique
in that they comprise a group of variables
with almost a constant mass obeying PLRs.
For instance in the cases of classical Cepheids and Mira variables
the mass increases with increasing period.
It is therefore interesting that our PLRs can be reproduced by
the simple scheme with a constant mass and
no need to adopt a mass-luminosity relation.
\begin{table*}
\begin{minipage}{120mm}
\caption{The relation between the adopted bolometric correction
($\alpha _\lambda, \beta _\lambda$) and the PLR
(slope $\mu _\lambda$ and zero-point $\eta _\lambda$).
Observational values for $JH\Ks$ are from the present paper,
and those for $V$ and $I$ filters are from \citet{Pritzl-2003}.
\label{tab:PLRslope}}
\begin{center}
\begin{tabular}{clrllll}
\hline
Filter & \multicolumn{1}{c}{$\alpha _\lambda$}
       & \multicolumn{1}{c}{$\beta _\lambda$}
       & \multicolumn{2}{c}{Equation (\ref{eq:Pul2})}
       & \multicolumn{2}{c}{Observation} \\
 & & & \multicolumn{1}{c}{$\mu _\lambda$} & \multicolumn{1}{c}{$\eta _\lambda$} &  \multicolumn{1}{c}{$\mu _\lambda$} & \multicolumn{1}{c}{$\eta _\lambda$} \\ 
\hline
$V$   & $+2.1\pm 1.0$ & $-7.06$ & $-1.85$ & $-1.94$
      & $-1.64\pm 0.05$ & $-1.92$ \\
$I$   & $-2.0\pm 0.5$ &   9.11  & $-2.08$ & $-2.77$
      & $-2.03\pm 0.03$ & $-2.80$ \\
$J$   & $-4.75\pm 0.1$ &  19.40  & $-2.23$ & $-3.34$
      & $-2.23\pm 0.07$ & $-3.54$ \\
$H$   & $-7.0\pm 0.1$ &  27.66  & $-2.36$ & $-3.75$
      & $-2.34\pm 0.06$ & $-3.94$ \\
$\Ks$ & $-7.2\pm 0.1$ &  28.47  & $-2.37$ & $-3.81$
      & $-2.41\pm 0.06$ & $-4.00$ \\
\hline
\end{tabular}
\end{center}
\end{minipage}
\end{table*}

\subsection{Comparison with T2Cs in the LMC}

\begin{figure}
\begin{center}
\includegraphics[clip,width=0.99\hsize]{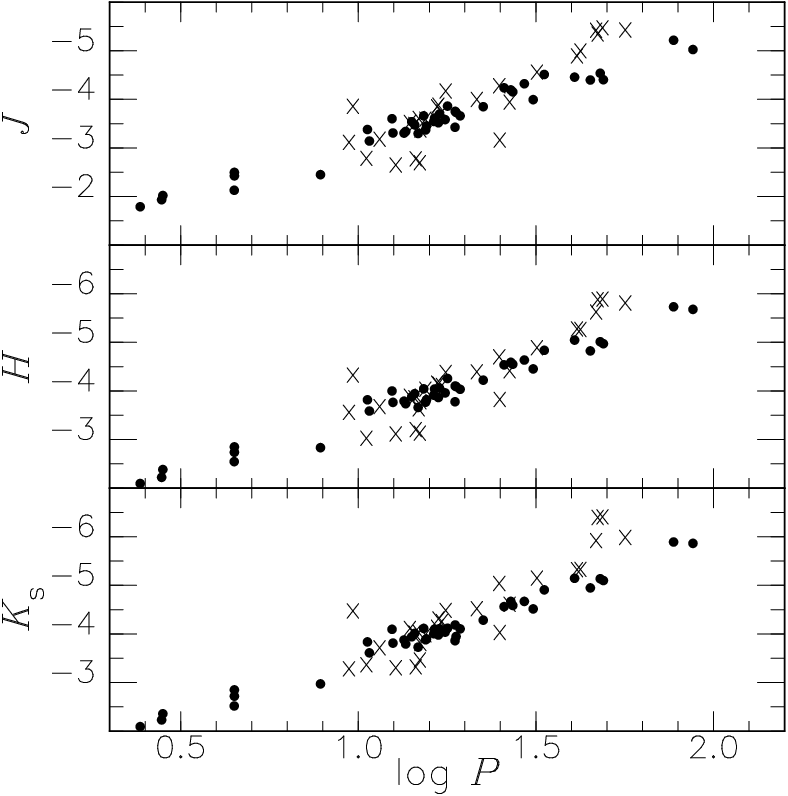}
\caption{
The PLR for the globular cluster sample
(filled circles) and the LMC candidates (crosses).
\label{fig:LMCcomp}}
\end{center}
\end{figure}

\citet{Alcock-1998} reported 33 candidate T2Cs with $8< P({\rm d}) <100$
based on the Massive Compact Halo Object (MACHO) data base in the LMC.
We searched for their near-infrared magnitudes in the 2MASS point-source
catalog\citep{Curti-2003}, and found 27 matches among 33 objects
(Table \ref{tab:LMC_Cep}).
Fig. \ref{fig:LMCcomp} shows the PLR in $JH\Ks$ for the T2Cs
in the LMC (crosses). Also plotted (filled circles)
are the T2Cs in globular clusters. The absolute magnitudes
for the LMC objects were obtained with an assumed distance modulus
of 18.50 mag.
There are some uncertainties of about $\pm 0.3$~mag in using
the 2MASS data because they are based on single-epoch observations.
However, it seems rather clear that longer period LMC variables ($P>40$d)
are brighter than the counterparts in globular clusters.
The LMC variables at shorter periods fit
the PLR of globular clusters within the uncertainties.
This feature is also seen in the $\log P$-$V$ diagram (fig. 9)
in \citet{Pritzl-2003}, but their conclusion is somewhat uncertain
because they do not have their own data for the cluster variables with
$P > 40$ d [compare the panels (a) and (d) of their fig. 9].

A mass difference could be one of the reasons 
for the LMC RV Tau stars being about one magnitude brighter than the
globular cluster PLRs.
Massive variables are expected to be brighter according to equation
(\ref{eq:Pul2}).
A difference of about 1 mag corresponds to
an increase in mass by a factor of  about 4.
Considering that the masses of the globular cluster variables are
about 0.5 -- 0.6 $M _\odot$, this would lead to
a mass larger than the Chandrasekhar limit for the
LMC RV Tau stars.
This is too large if these variables are post-AGB stars
which have already gone through their major mass-loss phase
(e.g. \citealt{Jura-1986}; \citealt{Pollard-1999}).
In that case some other parameter is necessary to explain the difference
in absolute magnitude.

\begin{figure}
\begin{center}
\includegraphics[clip,width=0.99\hsize]{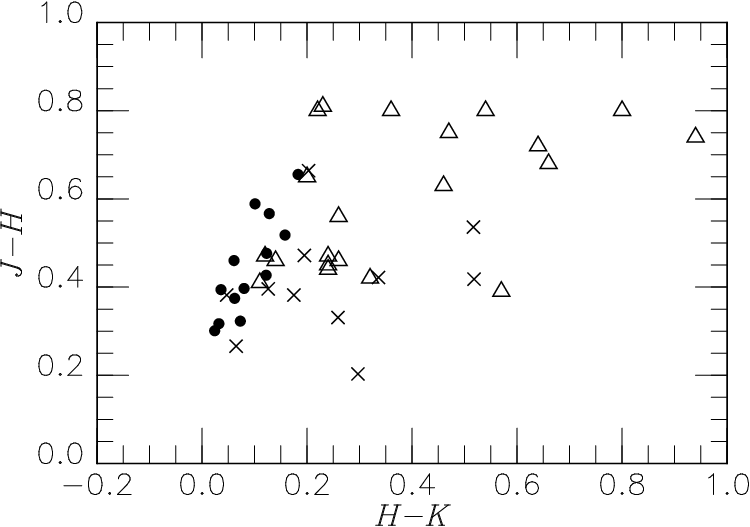}
\caption{
A colour-colour diagram for variables with $P>20$ d:
filled circles for globular cluster stars, crosses for the LMC stars
from \citet{Alcock-1998}
and triangles for galactic field stars\citep{LloydEvans-1985}.
The colours of the globular cluster stars and the LMC stars were corrected 
for the reddenings, while those of the galactic field stars were not
(see the text).
\label{fig:CCD_RV}}
\end{center}
\end{figure}

Fig. \ref{fig:CCD_RV} shows a colour-colour diagram
for the variables with $P> 20$ d in globular clusters
(filled circles) and the LMC (crosses) and also for the galactic field RV Tau
stars (triangles) taken from \citet{LloydEvans-1985}.
The colours of our cluster objects and the LMC objects were corrected 
for reddenings, while those of the galactic field objects were not
because \citet{LloydEvans-1985} did not give any estimate of the reddenings.
More than half of the galactic field objects have the galactic latitudes
of $|b|>5\degr$ so that the reddening effect on the colour is expected to
be not large ($E(H-K) < 0.1$). The large excesses of the $K-L$ colours
reported by \citet{LloydEvans-1985}, which should be smaller than $E(H-K)$
in the case of interstellar reddening,
also support that the objects are intrinsically red.
In Fig. \ref{fig:CCD_RV}, the globular cluster sample occupies a rather
limited region whilst many of the LMC and the local stars
spread to redder $H-\Ks$ colour.  
As mentioned in Section \ref{sec:Intro}, the classification of the 
longer period variables ($P> 20$ d) in globular clusters as RV Tau stars
is unclear.
One of the characteristics often seen in RV Tau stars
is an infrared excess caused by their circumstellar dust shells
\citep{Jura-1986}. The only globular cluster
RV Tau star which has been claimed to have an infrared excess is
NGC~6626 V17. \citet{Nook-1989} found an excess at 10 $\mu$m
in this star.
However, NGC~6626 V17 which has $(H-K)_0=0.12$ and $(J-H)_0=0.48$ 
lies with the other cluster stars 
in Fig. \ref{fig:CCD_RV},
and it also lies on our PLRs.
We conclude that RV Tau stars in the LMC belong to a different family
of variables from the T2Cs of the same periods in globular clusters.
Whether they define a PLR is not clear. 
\begin{table}
\begin{minipage}{86mm}
\begin{center}
\caption{2MASS magnitudes for candidate T2Cs in the LMC.
Star IDs are from \citet{Alcock-1998}.
\label{tab:LMC_Cep}}
\begin{tabular}{lrrrr}
\hline
Star ID & $P$ & $J$ & $H$ & $K_s$ \\
\hline
1.3812.61    &  9.387 & 15.453 & 14.970 & 15.222 \\
10.4040.38   &  9.622 & 14.635 & 14.153 & 13.998 \\
80.6469.135  & 10.509 & 15.749 & 15.483 & 15.125 \\
80.6590.137  & 11.442 & 15.347 & 14.824 & 14.771 \\
3.7332.39    & 12.704 & 15.868 & 15.380 & 15.178 \\
80.6475.2289 & 13.925 & 15.003 & 14.611 & 14.373 \\
81.9006.64   & 14.337 & 15.053 & 14.647 & 14.469 \\
47.2611.589  & 14.469 & 15.793 & 15.326 & 15.183 \\
19.4425.231  & 14.752 & 14.963 & 14.899 & 14.473 \\
2.5877.58    & 14.855 & 15.857 & 15.389 & 15.033 \\
1.3808.112   & 14.906 & 15.209 & 14.758 & 14.703 \\
14.8983.1894 & 15.391 & 14.988 & 14.500 & 14.526 \\
2.5025.39    & 16.602 & 14.723 & 14.369 & 14.368 \\
9.5117.58    & 16.747 & 14.697 & 14.394 & 14.199 \\
10.3680.18   & 17.127 & 14.836 & 14.424 & 14.254 \\
78.6338.24   & 17.560 & 14.354 & 14.120 & 14.002 \\
2.5026.30    & 21.486 & 14.571 & 14.136 & 13.984 \\
78.6698.38   & 24.848 & 14.281 & 13.823 & 13.463 \\
77.7069.213  & 24.935 & 15.364 & 14.676 & 14.457 \\
82.8041.17   & 26.594 & 14.584 & 14.088 & 13.877 \\
19.6394.19   & 31.716 & 14.011 & 13.641 & 13.356 \\
78.5856.2363 & 41.118 & 13.668 & 13.250 & 13.179 \\
81.8520.15   & 42.079 & 13.547 & 13.251 & 13.166 \\
82.8405.15   & 46.542 & 13.114 & 12.884 & 12.569 \\
81.9728.14   & 47.019 & 13.201 & 12.635 & 12.098 \\
79.5501.13   & 48.539 & 13.089 & 12.635 & 12.093 \\
47.2496.8    & 56.224 & 13.124 & 12.709 & 12.512 \\
\hline
\end{tabular}
\end{center}
\end{minipage}
\end{table}

\section{Summary}
\label{Summary}
We have shown from
our near-infrared observations
of T2Cs in globular clusters that they
define linear PLR at $JH\Ks$ with little
scatter.
There is no evidence for a
change of the slope at around $\log P=1$,
such as was suggested at $V$ in some  early papers.
An extrapolation of our infrared relation is shown to fit globular cluster
RR Lyr variables.
Both the slopes and the zero-points of the infrared PLRs can be
successfully reproduced by a simple application
of the pulsation equation at constant mass. The T2Cs and the RR Lyr
variables in clusters therefore seem to form an interesting family of
stars all with closely the same mass and showing a common PLR.

2MASS $JH\Ks$ magnitudes for W Vir stars and RV Tau stars
in the Large Magellanic Clouds show that,
within the uncertainties,
W Vir stars with $P<20$(d) obey the same PLR
as those for the globular cluster T2Cs. However,
RV Tau stars with $P>40$(d) 
are brighter than variables of the same periods in globular clusters.
The reason for this is unclear, but the distribution in
the $(H-\Ks)$-$(J-H)$ diagram also shows differences between the two group.
RV Tau stars in the LMC are generally redder than
those in globular clusters as are
RV Tau stars in the galactic field.

\section*{Acknowledgments}
We thank Dr Patricia Whitelock for her help on a part of observation.
The IRSF/SIRIUS project was initiated and supported by Nagoya University,
the National Astronomical Observatory of Japan and the University of Tokyo
in collaboration with the South African Astronomical Observatory
under a financial support of grant-in-aid for Scientific Research
on Priority Area (A) No. 10147207 and 10147214 
and the grant-in-aid for Scientific Research (C) No. 15540231
of the Ministry of Education,
Culture, Sports, Science and Technology of Japan.
One of the authors (NM) is financially supported
by the Japan Society for the Promotion of Science (JSPS)
through JSPS resarch fellowships for young scientists.

\bibliographystyle{mn2e}

\label{lastpage}

\end{document}